\def\RE{\mbox{$R_{\rm E}$}}
\def\DMM{\mbox{$M^{\rm E}_{\rm dm}$}}
\def\SM{\mbox{$M_{\rm *} (R_{\rm E})$}}
\def\TM{\mbox{$M_{\rm T} (R_{\rm E})$}}
\def\sL{\mbox{$\sigma_{\rm L}$}}
\def\sFJ{\mbox{$\sigma_{\rm FJ}$}}
\def\SMLR{\mbox{$M_*/L$}}
\def\reff{\mbox{$R_{\rm eff}$}}
\def\mlBulge{\mbox{$(M_*/L)_{\rm b}$}}
\def\mlDisk{\mbox{$(M_*/L)_{\rm d}$}}
\def\zl{\mbox{$z_{\rm l}$}}
\def\zs{\mbox{$z_{\rm S}$}}
\def\massDM{\mbox{$M_{\rm dm}$}}
\def\fracDM{\mbox{$f_{\rm dm}$}}

\newcommand{\gl}{OAC--GL~J1223-1239}

\documentclass[12pt,preprint]{aastex}

\usepackage{graphicx}

\shorttitle{Dark Matter in a lensing S0 galaxy at $z=0.47$}
\shortauthors{Covone et al.}

\begin{document}

\title{Gauging the dark matter fraction in a $L_*$ S0 galaxy at $z=0.47$
through gravitational lensing from deep HST/ACS imaging
\thanks{Based
on observations made with ESO Telescopes at the La Silla Observatories
and the NASA/ESA Hubble Space Telescope.}
}
\author{G. Covone,$^{1,2,3}$ 
M. Paolillo,$^{1,3}$
N.R. Napolitano,$^{2}$
M. Capaccioli,$^{1,4}$
G. Longo,$^{1}$
J.-P. Kneib,$^{5}$
E. Jullo, $^{5,6}$
J. Richard, $^{7}$
O. Khovanskaya,$^{8}$
M. Sazhin$^{8}$
N.A. Grogin,$^{9}$
and E. Schreier,$^{10}$}

\affil{$^{1}$Universit\`a di Napoli ``Federico II'', Naples, Italy\\
$^{2}$INAF--Osservatorio Astronomico di Capodimonte, Naples, Italy\\
$^{3}$INFN, Naples, Italy
$^{4}$INAF--VSTCeN, Naples, Italy\\
$^{5}$LAM-CNRS, OAMP,  Marseille, France \\  
$^{6}$ESO, Santiago, Chile \\
$^{7}$Department of Astrophysics, California Institute of Technology, Pasadena, USA\\
$^{8}$Sternberg Astronomical Institute, Moscow State University, Moscow, Russia\\
$^{9}$School of Earth and Space Exploration, Arizona State University,
Tempe, AZ 85287, USA\\
$^{10}$Associated Universities Inc., Washington, DC 20036, USA\\
}

\email{covone@na.infn.it}

\clearpage

\begin{abstract}

We analyze a new gravitational lens \gl, serendipitously found in a
deep I$_{814}-$band image of the Hubble Space Telescope (HST)
Advanced Camera for Surveys (ACS). The lens is a $L^*$, edge-on S0
galaxy at $z_{\rm l}=0.4656$. The gravitational arc has a radius of
$0''.42 \simeq 1.74 \, h^{-1} \, {\rm kpc}$. We have determined the
total mass and the dark matter (DM) fraction within the Einstein
radius \RE\ as a function of the lensed source redshift, which is
presently unknown. For $z_{\rm s}\sim 1.3$, which is in the middle
of the redshift range plausible for the source according to 
some external 
constraints, we find the central velocity dispersion to be
$\sim 180 \, {\rm km \, s}^{-1}$. With this value, close to that
obtained by means of the Faber-Jackson relation at the lens
redshift, we compute a 30\% DM fraction within $\RE\ $
(given the uncertainty in the source redshift,
the allowed range for the DM fraction
is 25-35 \% in our lensing model).
When compared with the galaxies in the local Universe, the lensing
galaxy \gl\ seems to fall in the transition regime between massive
DM dominated galaxies and lower-mass, DM deficient systems.

\end{abstract}


\section{Introduction}

Strong gravitational lensing (GL) is 
a valuable astrophysical tool to investigate the structure and
evolution of early-type galaxies up to a redshift of $\sim 1$
(e.g., Keeton et al.~1998; Rusin et al.~2003, Treu et al.~2005)
and to gauge the DM content at various galaxian scales (e.g., Treu
et al.~2006, Jiang \& Kochanek 2007). This tool offers the
advantage of constraining the total mass within the Einstein
radius independently of the dynamical status of the lensing
galaxy. The technique is challenging to apply to low-luminosity lenses
(LLLs), however, as the GL cross section is proportional to the fourth
power of the central velocity dispersion (e.g., Covone et
al.~2005). This is why LLLs appear far more rare than more massive
systems.  Their systematically smaller Einstein radii make strong
lensing arcs around them hard to find with optical surveys. For
instance, a lensing galaxy with velocity dispersion $\sL\sim 200
\, {\rm km \, s^{-1}}$ at $\zl \sim 0.5$, coupled to a source at
$\zs \sim 1.0$, produces an Einstein ring with radius 
$\theta_{\rm E} \sim 0''.5$. 
As a comparison, in a visual search for gravitational lenses 
in the COSMOS survey (Faure et al.~2008)\footnote{{\tt http://cosmosstronglensing.uni-hd.de/}}, 
no lensing galaxy was found with
central velocity dispersion smaller than $200~$km s$^{-1}$, 
out of a sample of 20 secure systems.

Although difficult to discover, LLLs are of great interest regarding
recent claims of dark-matter deficient $L_*$ galaxies in the
local Universe (see, e.g., Capaccioli et al.~2003, Romanowsky et
al.~2003, Napolitano et al.~2005, N+05 hereafter). Furthermore,
these systems occupy as region of the luminosity/mass distribution in
between boxy/slow-rotator systems and disky/fast-rotator systems (Nieto \& Bender
1989, Capaccioli et al. 1992). The dichotomy of these systems 
appears to involve also their DM properties, either at effective radii 
scales (Cappellari et al.~2006) or beyond (Capaccioli et al.~2003, N+05).

This paper analyses a $L_*$ edge-on S0 lensing galaxy at $z_{\rm
l}=0.4656\pm 0.0004$, hereafter named \gl, serendipitously
discovered in the HST/ACS follow-up imaging of a former candidate
lensing cosmic string (Sazhin et al.~2007). The mean radius of the
gravitational arc is $ \theta_{\rm E} = 0''.42$, corresponding to
a linear radius of $1.74 \, h^{-1} \, {\rm kpc}$, slightly larger
than the estimated value of the effective (half-light) radius
$\reff$ of the lens bulge (i.e. $\theta_{\rm E} \sim 1.15 \, \reff
$).  This compact arc offers the rare opportunity to gauge the mass
distribution within the effective radius of a $L_*$ galaxy at $z
\sim 0.5$. Note that, among the 20 strong GLs 
found in the 1.64 square degrees of the COSMOS survey by Faure et
al.~(2008),  selected in the range 
$ 0.2 < z < 1.0$ and median redshift 0.71, 
only one lens exhibits an arc with an angular radius smaller than $0''.40$.  

As the redshift of the lensed source is yet unknown, 
despite two spectroscopic runs (see Sect.~2),
the DM content
of the lens cannot be unambiguously determined. However, by
requiring that the lensing galaxy follows the Faber-Jackson 
relation (Faber \& Jackson~1976), and exploiting
the decoupled geometry of the luminous and the total mass, we can then
constrain the DM fraction within the Einstein radius.
Toward this end, in Sect.~3 we present a lensing model, and in Sect.~4 we
infer the luminous mass from the analysis
of the spectro-photometric data, in order to disentangle the lensing
contribution of the DM from that of the total mass.
In Sect.~\ref{solution}, we
discuss the best mix of dark and luminous mass which produces the
measured total mass as a function of the (unknown) source redshift,
and draw conclusions in Sect.~\ref{concl}.

Throughout the paper we will assume a cosmological model with
$\Omega_{\rm m}=0.27$, $\Omega_{\Lambda}=0.73$ and $h \equiv H_{0}
/ (100 \, {\rm km} \, {\rm s}^{-1}$\,Mpc$^{-1}) = 0.7$. At the
lens redshift, $1''$ corresponds to $4.15 \, h^{-1}$ kpc (Komatsu
et al.~2008). Magnitudes are in the AB system.

\section{The data}\label{data}

The gravitational lens \gl \, (RA$_{\rm J2000} = $ 12h~23m~32.65s;
Dec$_{\rm J2000} =  -12^{\rm o}~39'~40''.7$) falls in the
Osservatorio Astronomico di Capodimonte Deep Field (OACDF,
Alcal\'a et al. 2004), a survey performed with the 2.2m Wide Field
Imager (WFI) in three broad bands (B, V, R) and six intermediate
filters over the wavelength range 7730 - 9130 \AA. OACDF broadband
photometry  and coordinates of the lens are given in Table
\ref{table:oacdf}.
The spatial resolution ($FWHM\sim 1''$) of the best stacked OACDF
image (R band, limiting magnitude $R = 25.1$) was not sufficient
to resolve the gravitational arc, which was instead discovered
serendipitously by HST observations.

\begin{figure*}
\center
\includegraphics[width=0.5\textwidth, angle=0]{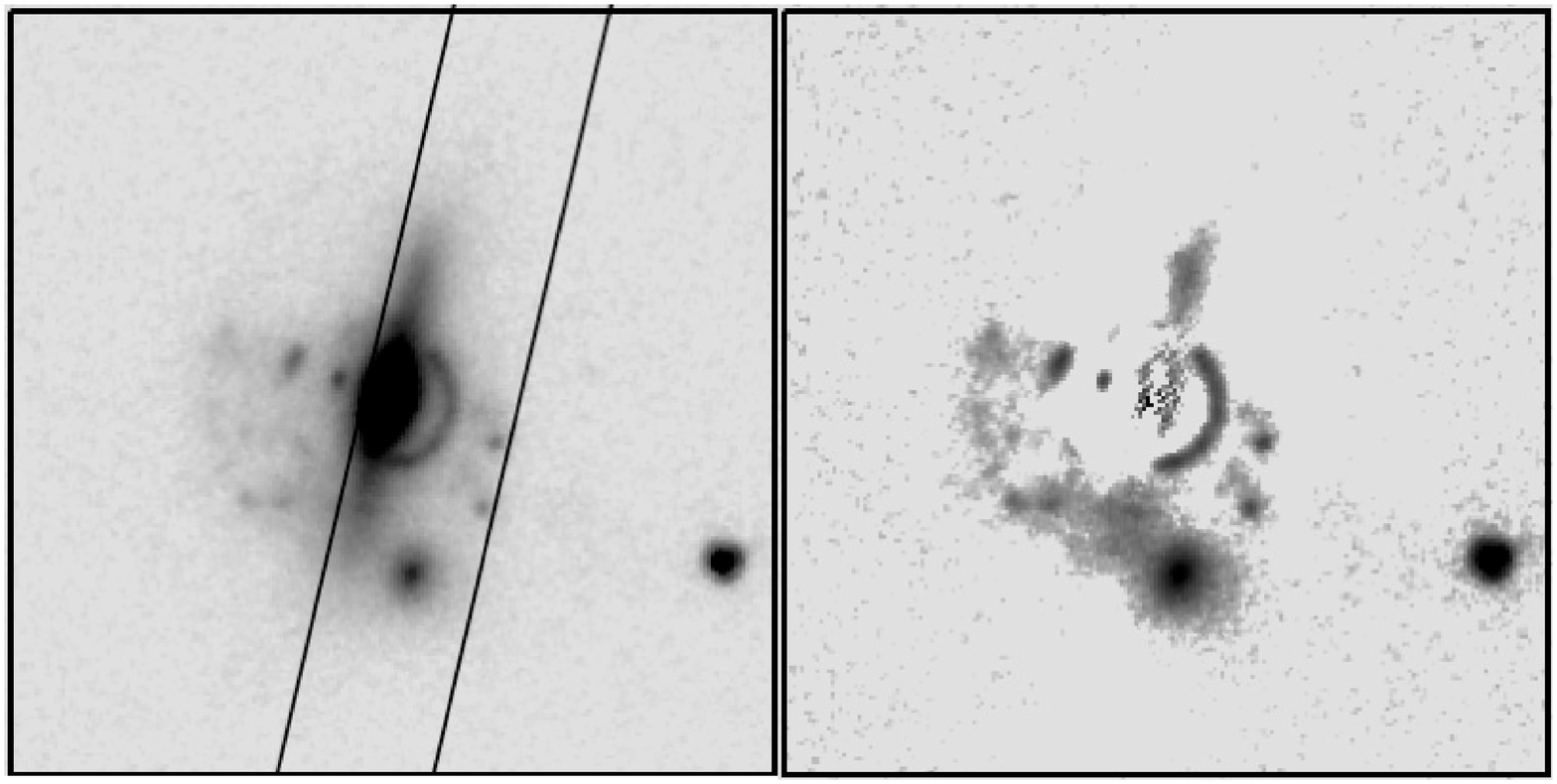}
\includegraphics[width=0.5\textwidth, angle=0]{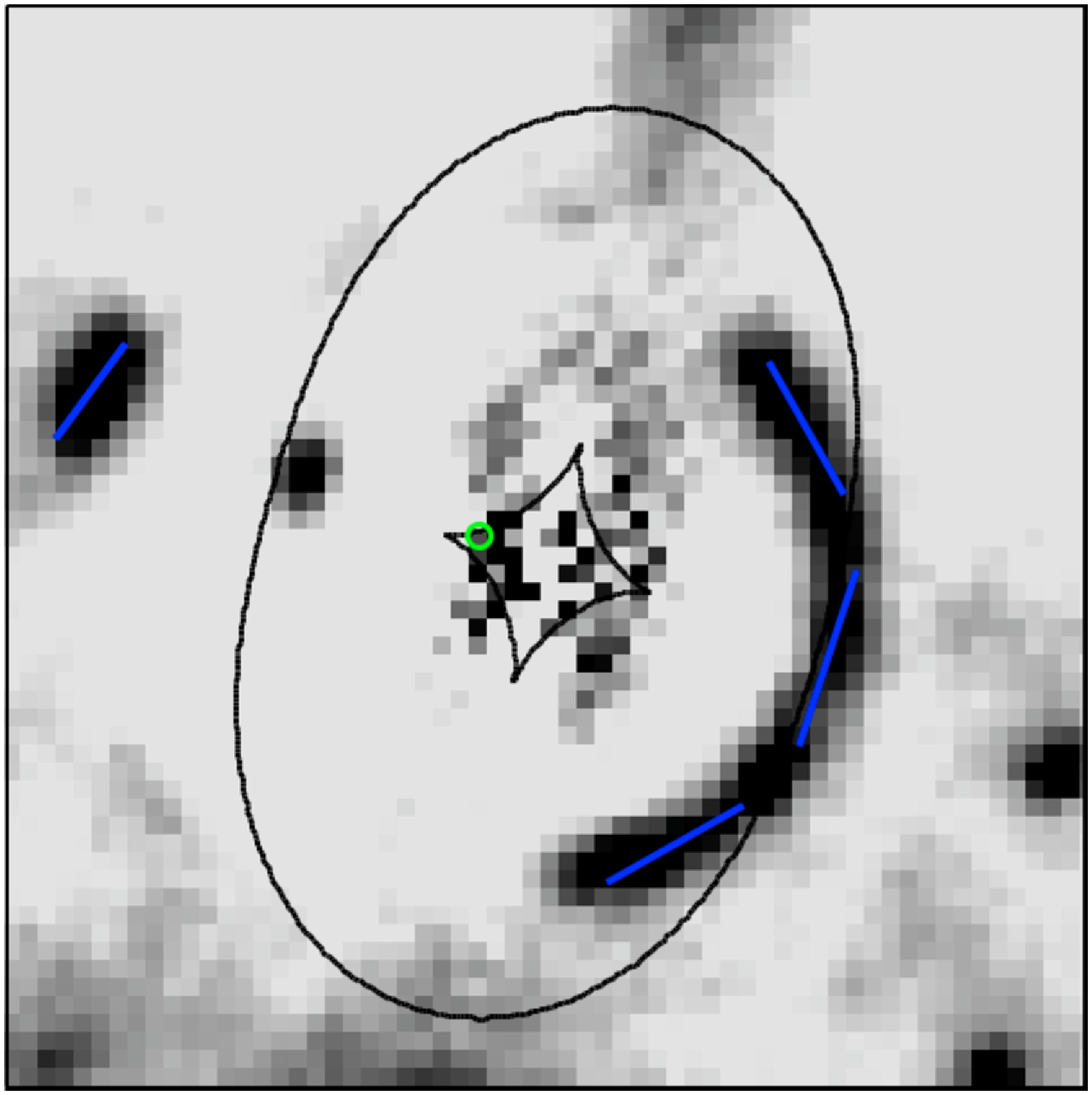}
\caption{Left panel: HST/ACS image of the lensing system in the I$_{814}$ filter;
the position of the slit used in the Keck/LRIS spectroscopic observations is shown;
field size is $5'' \times 5''$.
Middle panel: Same sky region, after subtracting the photometric model of the lensing
galaxy.
Right panel: the arc and counter-image positions (blue), 
caustics and critical line (black), 
predicted source location and size (green).
Field size is $1.5'' \times 1.5''$
}
\label{fig:one}
\end{figure*}

The latter are deep HST/ACS WFC follow-up images\footnote{Proposal
ID~10715, PI: M. Capaccioli. The same field was also observed with
HST/ACS by the program 10486, using filters F775W and F606W and
with shorter exposures (2409 and 5062~s, respectively). However, the
filter F775W covers very similar wavelength range to the I$_{814}$
image, offering no additional information, and in the F606W-band
data our target is a few arcsec out of the field-of-view.} collected
through the F814W filter ($I_{814}$ in the following) for a total
exposure time of 15~ks. A 1/3~pixel dither pattern (0\farcs05 per pixel) 
was adopted for sub-pixel sampling of the HST PSF and for cosmic ray
rejection. The stacked image, obtained by combining the 12 exposures
through the software Multidrizzle (Koekemoer et al. 2002), has
$0\farcs025$ per pixel sampling with spatial resolution of  $0''.10$ 
and 5$\sigma$ detection limit of $\sim 27.3$ mag arcsec$^{-2}$ (see 
details in Sazhin et al.~2007).
The HST/ACS image of the lensing system is shown in the left panel
of Fig.~\ref{fig:one}. In the right panel of Fig.~\ref{fig:one} we
show the residuals after subtracting the lens photometric model in
order to enhance the details of the lensed images. The model is
based on the surface photometry performed with the tool {\tt
galfit} (Peng et al.~2002). The best fit model requires two
photometric components: a $r^{1/4}$ (de Vaucouleurs 1948) bulge
combined with an exponential disk. The photometric parameters are
given in Table~\ref{table:phot}. Both the photometric model and
the spectroscopic data (discussed below) confirm that the lensing
galaxy is an edge-on S0.

The gravitational arc, subtending $\sim 150 \, {\rm degree}$, is
located at $0''.42$ from the lens centre,
perpendicular to the minor axis of the galaxy. A candidate
counter-image appears on the opposite side at
$0''.65 $. The nature of the other sources
around the galaxy is unclear: they could be part of the S0
galaxy and/or could be background objects, possibly linked with the strongly
lensed source.

\begin{table}
\caption{Broad-band photometry of the lens OAC--GL~J1223-1239}
\begin{center}
\label{table:oacdf}
\begin{tabular}{l l l}
\hline \hline
band     & total mag  \\
\hline
WFI B    & $ 22.74 \pm 0.05  $  \\
WFI V    & $ 21.86 \pm 0.04  $  \\
WFI R    & $ 20.85 \pm 0.03  $  \\
ACS F775 & $ 20.01 \pm 0.06 $  \\
ACS F814 & $ 20.04 \pm 0.05 $  \\
\hline
\hline
\end{tabular}
\end{center}
\end{table}

The spectroscopic information of the \gl\ rests on data collected
in two distinct observing runs. A low-resolution spectrum was
obtained in April 2000 with the ESO Multi-Mode Instrument at the
New Technology Telescope (NTT) in the multiobject spectroscopy
mode, within a survey of color-selected early-type galaxies at
intermediate redshift. Recently (while the paper was under
revision), a long-slit medium resolution spectrum has been secured
with the Low Resolution Imaging Spectrometer (LRIS) at the Keck I
Telescope.

\begin{figure}
\center
\includegraphics[width=0.5\textwidth, angle=0]{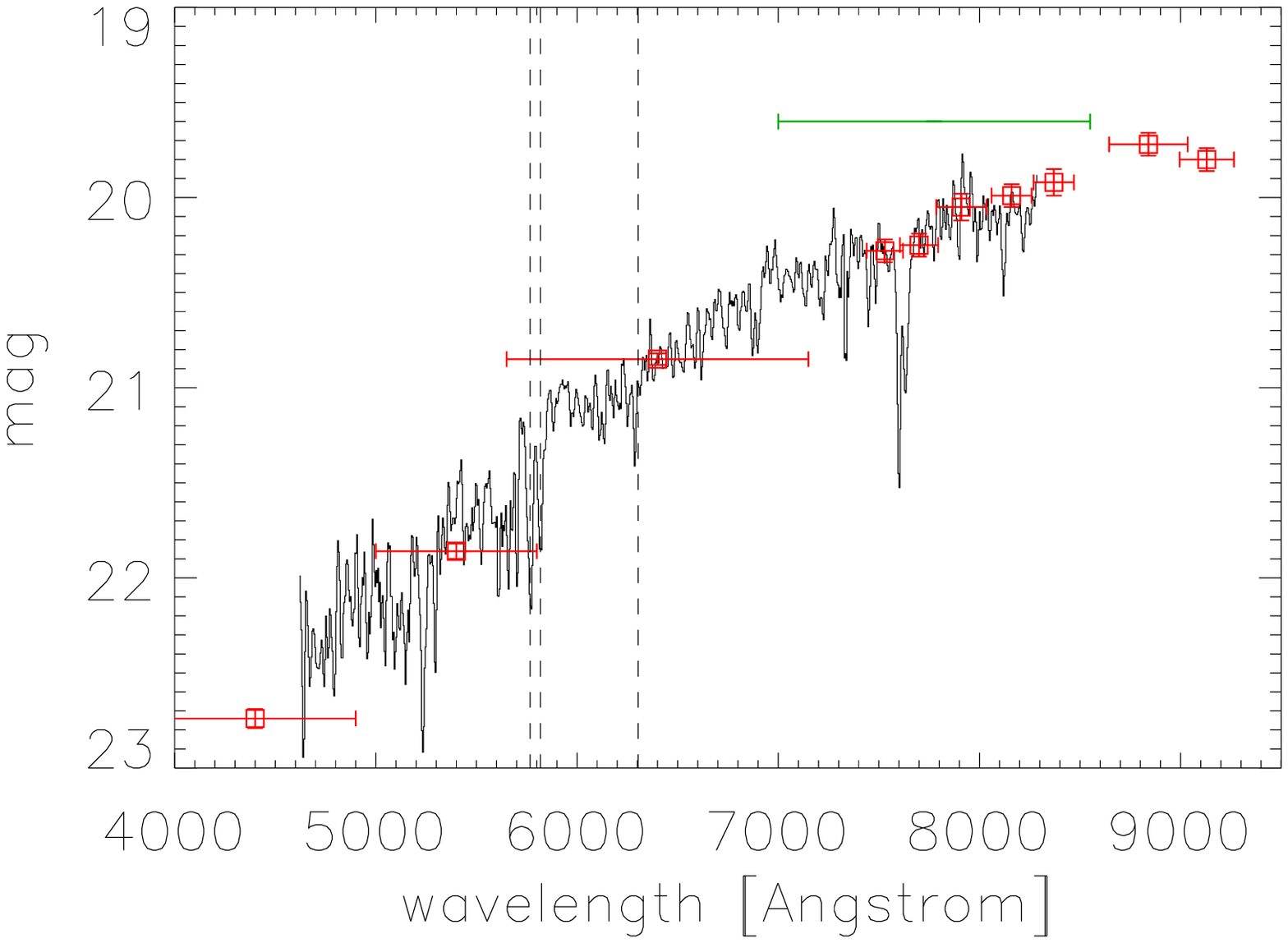}
\caption{Spectral energy distribution of the lensing galaxy:
the NTT spectrum (black) and
the total magnitudes (red points).
Dashed vertical lines show the positions
of the strongest spectral features
(Ca H and K bands and G band).
The horizontal green bar gives the
approximate comon wavelength range covered by the F775 and F814 filters.}
\label{fig:spec}
\end{figure}

The total exposure time of the NTT spectrum
was $3 \times 2400$ sec, with a slit of 1 arcsec and the grism
No.~3, yielding a dispersion of 2.3 \AA/pixel.
The spectral resolution
is $\simeq 10$ \AA ; the signal-to-noise about 8. Details of the
data reduction are given in Alcal\'a et al. (2004).
%
The NTT spectrum is shown in Fig.~\ref{fig:spec}, with the OACDF
broad- and narrow-band imaging photometry overplotted. Some
features typical of an early-type galaxy are apparent (the 4000
\AA \, break, the H and K Ca II absorption bands), thus confirming
the S0 classification and providing the redshift estimate
$\zl=0.466\pm0.005$. No clear feature from the lensed source was
observed because of the low S/N: the average surface
brightness of the arc, $\mu_{\rm 814} = 22.3 \, {\rm mag \,
arcsec}^{-2}$, is too faint to provide identifiable spectral
features other than strong emission lines.

In order to measure the redshift of the gravitational arc, on May
9th 2008 we obtained a spectrum of \gl\ with an exposure time of
$3\times1200$ sec using the 300 lines/mm grating of Keck/LRIS and a slit
$1''$ wide centered on the arc and aligned along the disk axis.

The blue side of the instrument used a 300 lines/mm grating blazed
at 5000 \AA ; the red side used a 600 lines/mm 
grating blazed at 7500 \AA.  A dichroic at 6800\AA   
allows a single exposure to sample the wavelength range
3500-9400 \AA.
The spectral resolution in the blue and red parts of the spectrum
was 8.5 and 3.8 \AA, respectively.  Despite the
good seeing conditions ($\sim\ 0''.7$), the spectrum does not
allow us to disentangle the arc from the lens.  Also, no clear
feature from the arc is apparent, owing possibly to the abundance
of atmospheric emission toward the red end of the spectrum. The superior
spectral resolution of Keck/LRIS gives a more precise 
redshift for the lensing galaxy of $\zl = 0.4656 \pm 0.0004$.

The rest-frame absolute magnitude of the lens galaxy, derived from
the broad-band photometry (see Table \ref{table:oacdf}) assuming a
negligible extinction\footnote{${\rm E(B-V)}$ is below 0.03 all
over the OACDF; see Burstein \& Heiles (1982).} is M$_B = -20.7
\pm 0.1$. Thus \gl\ is an intermediate luminosity
galaxy (in the sense of the Schechter luminosity function's
parameter $L_*$) since, at $z \sim 0.45$, the absolute magnitude
of $L_*$ galaxies is $M_B^* = -20.78 \pm 0.17 $ (Bell et al.
2004).

The color of \gl, ($V-I) = 1.81$, is close to that of early-type
galaxies in clusters at the same redshift. For instance, in the
sample of objects at $z \sim 0.47$ within the ESO Distant Cluster
Survey (De Lucia et al.~2007), galaxies at $I \sim 20$ on the red
sequence have $(V-I) \simeq 1.65$.

\begin{table}
\caption{Photometric properties of the lensing galaxy, as observed in the band $I_{814}$.}
\begin{center}
\label{table:phot}
\begin{tabular}{l  l   l  l  l  l }
\hline
\hline
      & mag    & $q$ & PA    &  $\reff$ & $\reff$  \\
      &        &     & [deg] &          & [kpc] \\ \hline \\
bulge  & 20.45  &   0.35 &   -13.6 &  $0''.37$    & 1.54     \\
disk   & 21.59  &   0.25 &   -8.4  &  $0''.77$    &  3.16    \\
\hline
\end{tabular}
\tablenotetext{}{$q$ is the ratio of the two galaxy axes.}
\end{center}
\end{table}

\section{The lensing model and the total mass within \RE}\label{lensing}

The mass distribution of the lens component of \gl, which must
reproduce the morphology of the arc and of the counter-image,
was modeled by a singular isothermal ellipse (SIE) plus a
contribution from an external shear $\gamma$ to take into account
the effects of the galaxies close to the line-of-sight. The $\sim
r^{-2}$ total mass density profile of the our model is motivated
by several statistical studies of lens samples. For
instance, Koopmans et al.~(2006) find that the slope of the total
mass profile is isothermal at any redshift in the sample of SLACS
lensing galaxies. The parameters of the model were fit by means 
of the {\tt lenstool} software\footnote{Publicly available at {\tt
www.oamp.fr/cosmology/lenstool/}.} (Kneib 1993), based upon a
Bayesian Monte Carlo Markov Chain optimization method (see Jullo
et al.~2007 for details). The critical lines of the best SIE model
are drawn in Fig.~\ref{fig:one}, and the parameters listed in
Table~\ref{table:model}.

\begin{table}
\begin{center}
\caption{Parameters of the best gravitational lensing model}
\vspace{0.2cm}
\label{table:model}
\begin{tabular}{l l l }
\hline
\hline
$\chi^2 / {\rm n.d.f.}$                & 1.28                     &    \\
axis ratio                             & 0.55 $\pm$ 0.06          & \\
P.A. [deg]                             & $-26 \pm 5$              &  \\
external shear $\gamma$                & $0.080 \pm 0.026$        &   \\
$\theta_{\gamma}$ (deg)                & $ -72 \pm 20$            &  \\
magnification                          & $\sim 12$                &   \\
\sL\ ($z_s$=0.9) [km \, s$^{-1}$]        & 204  $\pm 3 $ &\\
\sL\ ($z_s$=1.6) [km \, s$^{-1}$]        & 170  $\pm 3 $ &\\
\hline
\end{tabular}
\end{center}
\end{table}

\begin{figure}
\center
\includegraphics[width=0.5\textwidth, angle=0]{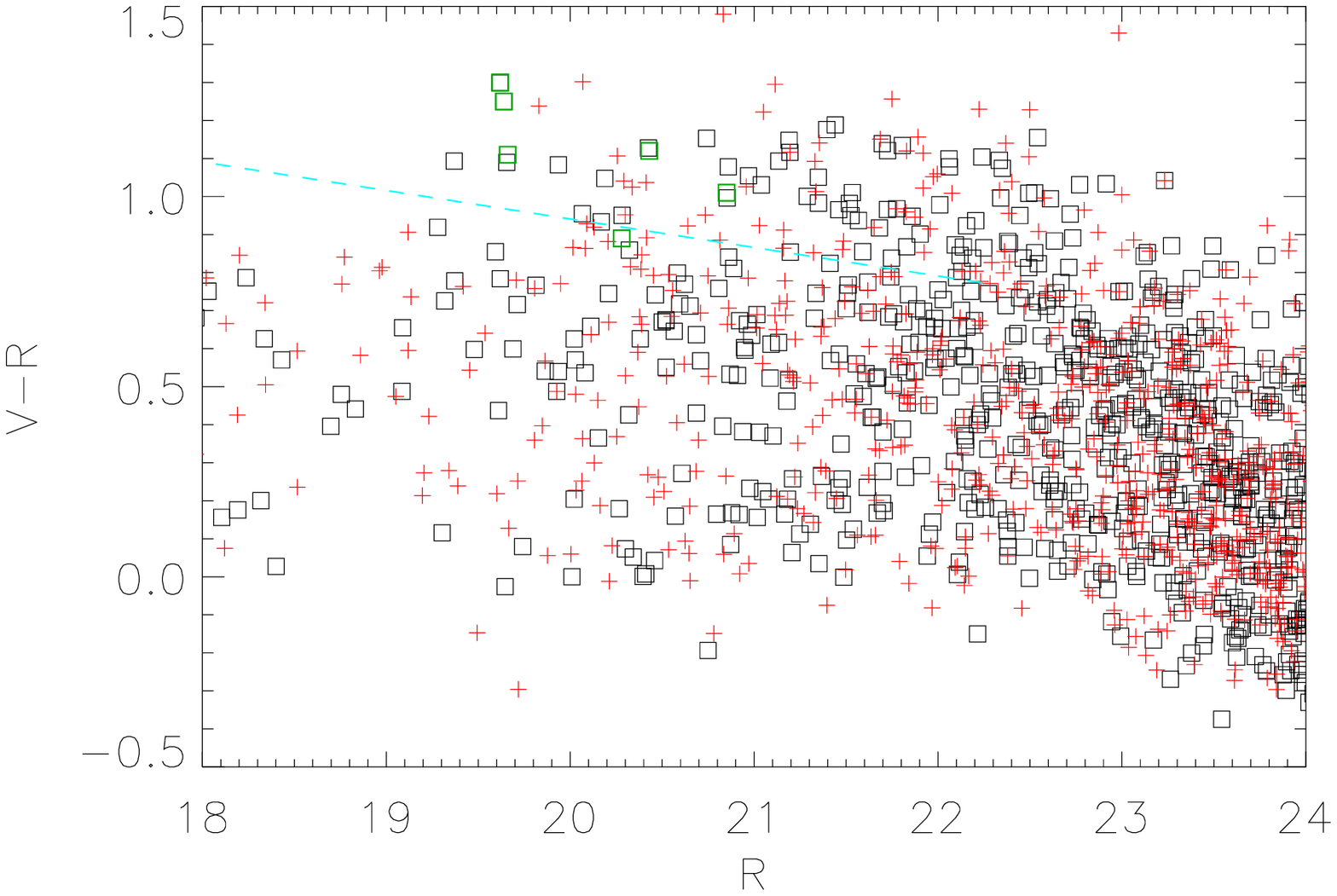}
\caption{Color-magnitude diagram of the galaxies
within 1 Mpc of the gravitational lens (black boxes),
compared with the sources from the whole OACDF survey
(red cross).
Galaxies at redshift $z = 0.46 \pm 0.01$
in the same spatial region are shown as
green boxes.
The overplotted line is the fitted red sequence
in galaxy clusters at $z\sim0.45$ observed by
De Lucia et al. (2007).}
\label{fig:cm}
\end{figure}

The best lensing model includes some external shear, possibly due
to a local galaxy overdensity.
The presence of a group of galaxies at $\zl$ is supported by both
photometric and spectroscopic data. A color-magnitude diagram of
the galaxies within 4~arcmin (i.e., a projected radius of $\sim
1$~Mpc) from the lensing galaxy, compared with the distribution of
galaxies from the whole OACDF survey (see Fig.~\ref{fig:cm}),
shows a dozen $L>L_* $ galaxies located along a red sequence
close to the one observed at $z \sim 0.45$ by De Lucia et
al.~(2007). Six galaxies have spectroscopic redshifts in the range
$z = 0.46 \pm 0.01$, including the \gl\ and a system of two bright
ellipticals (CSL1, $z = 0.463$, Paolillo et al.~2008) located at $53''$
(220~kpc $\, h^{-1}$) from the lensing galaxy.
Note that the spectroscopic survey of early-type galaxies in the
OACDF was not spatially complete and did not cover the whole
survey field. In particular, only $\sim 1/3$ of the region within 4.0
arcmin from the lens was covered by the masks.

The lensing model exhibits a small difference, $ \Delta \theta =
-13^{\rm o}  \pm 5^{\rm o}$, in the orientation of the total mass model (i.e.,
dark and luminous matter) with respect to the light distribution
(see Table~\ref{table:phot} and \ref{table:model}), which does not
disappear by forcing a stronger external shear.
This result is marginally consistent with Koopmans et al.~(2006)
who find, for lenses with velocity dispersion $ > 200 ~{\rm km \,
s}^{-1}$, that the total mass is aligned with the light to within
$10^{o}$ (see also Kochanek~2002). Furthermore, the total mass
distribution is more circularized ($q_{\rm SIE} = 0.55$) than the
light ($q_{\rm star}=0.35$).

This modeling points towards a geometry of the total mass
different from that of the stellar mass, which in turn suggests
the presence, within \RE, of a dark matter component with a different spatial
distribution than the light. 
By an iterative double-component
approach with a spherical DM halo coupled to a flattened ($q_{\rm
star}=0.35$) stellar bulge, we find that, within \RE, a dark
matter mass \DMM\ of the order of 25\% of the total mass \TM\ is
sufficient to account for the ratio $q_{\rm SIE}/q_{\rm star}$ and
to reproduce the observed image configuration.
A larger DM fraction ($\sim 40$\%) is found for a non-spherical ($q
\sim 0.8$) dark halo, in agreement with N-body simulations
at this mass scale (Bullock et al.~2001) as well as weak lensing measurements
(Hoekstra et al. 2004). It reduces to $\sim 35$\%
by forcing the halo to match the observed tilt.\\
In conclusion, this analysis constrains the dark mass fraction
within the Einstein radius to 25-35\% of \TM.

\begin{figure}
\center
\includegraphics[width=0.5\textwidth, angle=0]{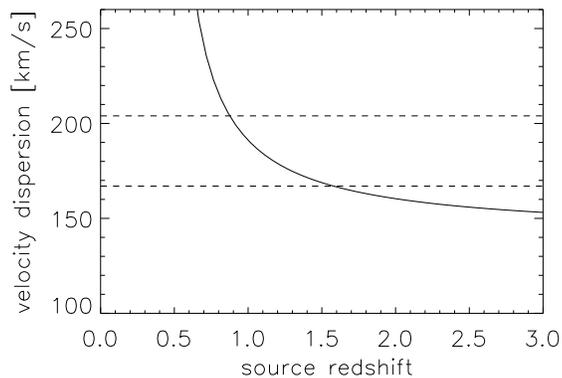}
\caption{Velocity dispersion of the best SIE lens model as a function of the source redshift.
Dashed lines correspond to the two FJ estimates (see text for details).}
\label{fig:sigma1}
\end{figure}

Our model additionally provides the central velocity dispersion of the lensing
galaxy as a function of the redshifts of the source and of the
lens:
\begin{equation}
\sL =  144 \, \times \, \sqrt{\frac{D_{\rm s} \, (z_{\rm
s})}{D_{\rm ls} \, (z_{\rm l}, z_{\rm s})} \, } \, {\rm km} \, {\rm s}^{-1} \, ,
\end{equation}
where $D_{\rm s}$ and $\, D_{\rm ls}$ are the angular diameter
distances to the source and from the lens to the source
respectively (see also Fig.~\ref{fig:sigma1}). The corresponding mass is:
\begin{equation}
\TM = 4.7 \, \times 10^{10} \, \frac{D_{\rm s} \, (z_{\rm
s})}{D_{\rm ls} \, (z_{\rm l}, z_{\rm s})} \, {\rm M}_{\odot}.
\end{equation}
For instance, with a source redshift ($z_{\rm s}=1.6$) corresponding
to the maximum strong lensing cross section of a lens at
$\zl=0.466$, one obtains $ \sL =  170 \, {\rm km \, s}^{-1}$ (see
Table~\ref{table:model}) and $\TM= 7.0\times 10^{10} M_\odot$.

We can compare the trend of \sL\ with the value of the velocity
dispersion \sFJ\ derived from the Faber \& Jackson relation (1976; FJ)
applied to \gl,
\begin{equation}
\frac{ L}{L_*} = \, \left(\frac{\sFJ}{\sigma_*}\right)^{\gamma} \,
\times  10^{\gamma_{\rm ev} z}
\end{equation}
corrected for galaxy passive evolution. Following Rusin et
al.~(2003), we  adopt a B-band FJ slope $\gamma = 3.29 $,
$\sigma_* = 225 \, {\rm km/s}$, and a passive evolution term
$\gamma_{\rm ev} = -0.41$. The result $\sFJ = 204 \pm 28 \, {\rm
km \, s}^{-1}$ is larger than the value obtained from the lens
model with $z_{\rm s}=1.6$, and requires a source redshift as low
as $z_{\rm s}=0.9$, implying $\TM= 10.1\times 10^{10}M_\odot$.

We note, however, that a large uncertainty still remains on the
value of the critical parameter $\sigma_*$. For instance, Davis et
al.~(2003) argue in favor of a significantly lower value,
$\sigma_* \sim 185 \pm 15 \, {\rm km \, s}^{-1}$, by which we
obtain $\sFJ = 167 \pm 23 \, {\rm km \, s}^{-1}$, identical to the
value of \sL\ with $z_{\rm s}=1.6$.

In summary, the FJ relation constrains $z_{\rm s}$ in the wide
range 0.9  - 1.6, equivalent to a $\sim$30\% variation of the
total mass within the \RE. This is three times larger than typical
mass uncertainties at these radii (Kochanek et al.~2000) and does
not independently allow to draw any firm conclusion on the 
DM content of the system.

However, besides the geometrical constraint discussed above
($\DMM\simeq 25-35\% \, \TM$) and that given by the Faber-Jackson
relation ($144 < \sL < 232 \, {\rm km \, s}^{-1}$, including errors),
we have still another card to play: we may estimate the stellar
mass \SM\ within \RE\ from the spectro-photometric data.

The Keck/LRIS spectroscopic data allow us to measure the average velocity
dispersion $\overline{\sigma}$ of the lensing galaxy over the region
covered by the slit, which is $\sim 3 \reff$ wide and slightly
off-center with respect to the galaxy nucleus (see Fig.~1).
The fit of the $\lambda5892$ absorption line profile gives
$\overline{\sigma}=150 \pm 11 \, {\rm km s}^{-1}$.
By modeling the galaxy as in Napolitano et al. (2008, in preparation), this figure
returns an estimate of the central velocity dispersion in agreement
with the lower value predicted by the FJ relation, i.e. 
$\sFJ = 170  \pm 15 \, {\rm km s}^{-1}$.
Mindful of the uncertainty intrinsic to this extrapolation, in
the following analysis we will consider the whole range of values allowed
by the FJ relation.


\section{The stellar mass content within $R_{\rm E}$}
The stellar mass-to-light ratio, \SMLR, has been derived by fitting
the low-resolution spectrum (Fig.~2) with the library of synthetic
spectra by Bruzual \& Charlot (2003). Since the choice of the
initial mass function (IMF) is a critical point, we shall explore
both Salpeter (1955) and Chabrier (2003) IMFs. 
A Salpeter IMF gives
a global stellar mass-to-light ratio $M_* / L  = 4.0 \, M_{\odot} /
L_{\odot}$, in the rest-frame B-band, with an age of 8 Gyr and a
sub-solar metallicity ($Z = 0.004 \, Z_{\odot}$). If \SMLR\ is
uniform, from the total luminosity $L_{\rm B} = 2.7 \times 10^{10}
\, L_{\odot}$ one obtains a total stellar mass $ M_* = 1.1 \times
10^{11} $  M$_{\odot}$.

In order to compute $\SM$, the stellar mass within \RE, it is
prudent to disentangle the contributions of the two components,
disk and bulge, which have different light profiles, isophotal
geometry (see Table~2), and possibly mass-to-light ratios.
Evolutionary synthesis models of template S0 galaxies of age $\sim
8$ Gyr predict a bulge mass-to-light ratio, $(M_*/L)_{\rm b}$, in
excess of up to 1.5 times that of the disk, $(M_*/L)_{\rm d}$
(Buzzoni 2005).  We  use this upper limit to separate the
contribution of the two components.

By definition,
\begin{equation}
M_*= \mlBulge \times L_{\rm bulge}+ \mlDisk  \times L_{\rm disk}
\label{eq:ml}
\end{equation}
where $ L_{\rm bulge} = 1.9 \times 10^{10} L_{\odot} $ and $L_{\rm
disk} = 0.8 \times 10^{10} L_{\odot}$ from the measured
bulge-to-disk ratio $B/D \sim 2.2$ (Table~2). Solving
Eq.~(\ref{eq:ml}) for a Salpter IMF, we obtain $\mlDisk = 2.9 \,
M_{\odot} / L_{\odot}$ and $\mlBulge = 1.5 \times M/L_{\rm d} = 4.4
\, M_{\odot} / L_{\odot}$. Thus, the bulge and disk masses within
$R_{\rm E}$ are $M^E_{\rm b}=5.4 \times 10^{10} M_{\odot}$ and
$M^E_{\rm d}=0.3 \times 10^{10} M_{\odot}$ respectively, with a
total stellar mass $\SM=5.7 \times 10^{10} \, M_{\odot}$.

A simpler model with a single value of $M_*/L$ gives a total mass
(bulge+disk) within $R_{\rm E}$ of $5.2 \times 10^{10} \, M_{\odot}$
(Table~4).

Repeating the calculations with a Chabrier IMF, the stellar \SMLR\
ratio is $\sim 1.8$ smaller than with Salpeter ($\SMLR = 2.2 \,
M_{\odot} / L_{\odot}$ for the single population model, and
$M/L_{\rm b}= 2.6$, $M/L_{\rm d}=1.6\, M_{\odot}/L_{\odot}$ for the
two population model), with the same age and metallicity assumed
above. The stellar masses of bulge and disk are 1.8 times smaller,
accordingly (Table~4).

\begin{table}
\begin{center}
\caption{Stellar masses at the Einstein radius.\label{table:Mstar}}
\begin{tabular}{llllll}
\tableline
\tableline
\multicolumn{5}{c}{Salpeter  IMF} \\
\tableline
model &  $M/L_{\rm b}$  & $M_{\rm b}$&  $M/L_{\rm b}$ & $M_{\rm d}$
& $M_* (R_{\rm E})$  \\
\tableline
$M/L_{\rm b} = 1.5 \times M/L_{\rm d}$  &  4.4 &  5.4  &  2.9  &  0.3  & 5.7 \\
$M/L_{\rm b}=M/L_{\rm d}$ &  4.0 &  4.8  &  4.0  &  0.4  & 5.2  \\
\tableline
\\
\multicolumn{5}{c}{Chabrier IMF}  \\
\tableline
model     & $M/L_{\rm b}$  & $M_{\rm b}$ &  $M/L_{\rm b}$ & $M_{\rm d}$
& $M_* (R_{\rm E})$ \\
\tableline
$M/L_{\rm b} = 1.5 \times M/L_{\rm d}$  &  2.6 &  3.0  &  1.6  &  0.2  & 3.2\\
$M/L_{\rm b}=M/L_{\rm d}$ &  2.2 &  2.7  &  2.2  &  0.2  & 2.9   \\
\tableline
\end{tabular}
\tablenotetext{}{Mass-to-light ratio are given in units of
$M_{\odot}/L_{\odot}$ and stellar masses in units of $10^{10} \, M_{\odot}$.}
\end{center}
\end{table}

\section{Comparing lensing and stellar masses} \label{solution}
Since the quantity $\TM$ remains parametrised with the source
redshift (see Sect.~3), we now use the $\SM$ values derived in the
previous Section to infer the range for $z_{\rm s}$ compatible with
the other constraints. Figure~\ref{fig:tre} shows the trend for the
projected DM fraction, $\fracDM (R_{\rm E})=\DMM/\TM$, as a function
of the $z_{\rm s}$, for both Salpeter and Chabrier IMFs, together
with the constraints by the FJ relation and by the geometry of the
system (green region). The redshift range allowed by all the
constraints is $z_{\rm s}= 1.2-1.5$ in the case of a single
population model and $z_s=1.1-1.3$, in the case of a composite
disk+bulge population.  The Chabrier IMF is
highly inconsistent with the given limits.

In conclusion, considering a DM fraction (within \RE) in the range
25-35\%, the most likely solutions correspond to the redshift range
$z_{\rm s}  = 1.3 \pm 0.2$ and favor the Salpeter IMF.

\subsection{Interpreting the favored lens model in the $\Lambda$CDM framework}
%
The total $M/L$ within \RE\
varies from $4.7 \, (z_s=1.5)$ to $6.1 \, (z_s=1.1)$,
with an average value of
$M/L(R_{\rm E})=5.4$--5.5 for $z_{\rm s}=1.3$.
As shown in Sect.~4, considering just the baryons contribution, $M/L_*=4.0$--4.4, allowing
for the two-population model (Table~2);
that is, the total $M/L$ is a factor $\sim 1.2-1.5$
larger than the stellar $M/L$ over the whole redshift range.
These variations
can be interpreted with the toy model from
N+05 to infer the global properties of the virial DM content
of the lensing system in the $\Lambda$CDM cosmology
(see also Ferreras et al.~2005 for an application to
a sample of lensing galaxies).

We consider a multi-component galaxy model with bulge stars
distributed following a Hernquist (1990) profile with effective
radius and total luminosity as from the first row of Table~2, disk
stars following an exponential profile, with parameters as in the
second row of Table~2, and a spherical DM halo with a Navarro, Frenk
\& White (NFW; 1998) density profile. As in N+05 and Bullock et
al.~(2001), we adopt the following concentration-mass relation for
the DM halo:
\begin{equation}
c_{\rm dm}(M_{\rm dm})\simeq \frac{1}{1+z} 17.1
\left( \frac{M_{\rm dm}}{10^{11} M_{\odot} \, h^{-1} }\right )^{-0.125} \, ,
\label{cMvir}
\end{equation}
where the concentration parameter is $c_{\rm dm} \equiv r_{\rm
vir}/r_s$, with $r_{\rm s}$ the characteristic scale of the NFW
profile, and $M_{\rm dm}$ is the total dark halo mass at the virial
radius $r_{\rm vir}$. The cumulative mass profile for a NFW dark
halo is
\begin{equation}
M_{\rm dm}(r) = M_{\rm dm} \frac{A(r/r_s)}{A(c_{\rm dm})} \, ,
\label{MNFW}
\end{equation}
where
\begin{equation}
A(x)=\ln (1+x)-\frac{x}{1+x} \, .
\end{equation}

We can now estimate the ratio $f_{\rm vir} \, \equiv \massDM
/M_{*}$, computing the NFW halo mass from Eq.~(\ref{cMvir}) and
Eq.~(\ref{MNFW}), and the stellar mass associated to the Hernquist
model by using the photometric parameters of bulge and disk (see
Table~\ref{table:phot}).
Imposing that the modeled $M/L$ is a factor $\sim 1.2-1.5$ larger
than the stellar $M/L$ within $R_{\rm E} (\sim 1.15 \,
\reff$)\footnote{This corresponds to a logarithmic $M/L$ gradient,
$\nabla \Upsilon\equiv \frac{\reff \Delta \Upsilon}{\Upsilon_{\rm
in} \Delta R}\sim0.2-0.3$ (see N+05) which is typical of local
low-$\nabla \Upsilon$, i.e. low-DM systems.}, we find $f_{\rm vir}
\, \sim 6 \pm 4$, for the whole galaxy. Here the main uncertainty
comes from the choice of the stellar $M/L$, rather than from the
halo concentration from the adopted c$_{\rm dm}-$M$_{\rm dm}$ (see,
e.g., Bullock et al.~2001) and the total $M/L(R_{\rm E})$ estimates.
In Fig.~\ref{fig:tre} the projected DM fraction is shown together
with the corresponding values of $f_{\rm vir}$ as a function of the
source redshift. If we consider the whole allowed range of $z_{\rm
s}$ and the effect of the choice of the stellar $M/L$ on the DM
fractions, we see that the largest $f_{\rm vir}$ value is $\sim$20
for $z_{\rm s}$=1.1 (i.e., the lowest redshift allowed), while
$f_{\rm vir}<20$ for $z>1.1$, and $f_{\rm vir}<10$ for $z_{\rm
s}>1.2$.

These values are consistent with the low global DM fraction regime
at $z=0$ discussed in N+05. These authors found a transition mass
around $M_{\rm *}=1.6\times 10^{11} M_\odot$ between more massive DM
dominated systems and low-mass, DM poorer systems. Incidentally, \gl
\, is exactly located in this transition regime.

\section{Discussion and conclusions}\label{concl}

\begin{figure}
\center
\includegraphics[width=0.5\textwidth, angle=0]{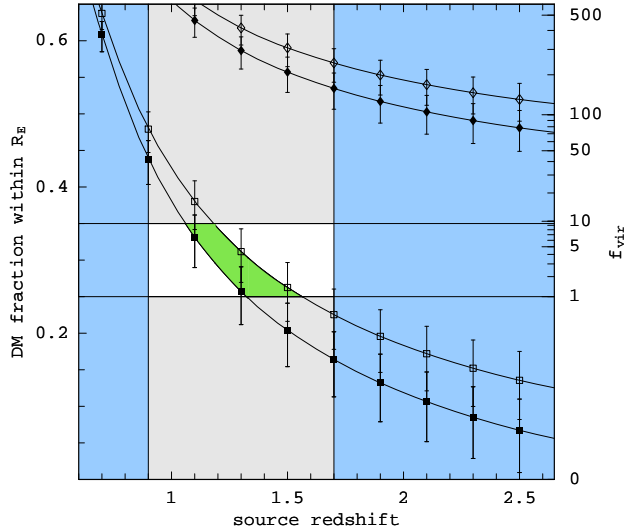}
\caption{DM fraction at $R_{\rm E}$ versus the (unknown) redshift
of the lensed source. On the right the corresponding (mean) $f_{\rm vir}$ values;
see text for details.
Empty boxes are the DM
fractions for the single bulge+disk population, full boxes for the
two population models,
assuming Salpeter IMF for the stellar $M/L$.
Diamonds represent the same models for the Chabrier IMF.
Grey denotes the regions excluded by the constraints on the $\fracDM$ from the lensing model geometry,
cyan are the regions excluded by the Faber-Jackson relation.
In green, the range of the parameter
space allowed by all the contraints, accounting also for the uncertainty associated
to the population models $M/L$.}
\label{fig:tre}
\end{figure}

\gl, a $\sim L_*$ S0 galaxy at redshift $z=0.466$, is
among the least massive lenses at high redshift known to date.
Although the source redshift is still not known,
the observed phenomenology unambiguously supports the strong-lensing 
interpretation.
The lensed source produces a bright arc (see Fig.~1)
with radius  $0''.42$ ($\sim 1.15 \, \reff$ of the lens bulge).
A candidate counter-image is found at $0''.65$ from the lens center.

The total mass distribution, modeled by a singular isothermal
ellipse, is found to be rounder than the light at the Einstein
radius, with a ratio $ q_{\rm SIE} / q_{*} = 1.84$, similar to the
value found for galaxies with velocity dispersions lower than $\sim
200 \, {\rm km \, s}^{-1}$.

We constrain the redshift of the source by
coupling the value of the ratio
$(\massDM /M_*) (\RE)$, obtained via the lens model,
with the lens total mass within $R_{\rm E}$, 
once the  corresponding stellar mass \SM\
is evaluated by an assumption
on the baryonic component
and corresponding stellar $M/L$.
We obtain $z_{\rm s} \sim 1.3 \pm 0.2$ with a Salpeter IMF,
the Chabrier IMF being ruled out.
In this case the velocity dispersion
from the lens model
(a singular isothermal ellipse with an external shear)
is $\sL\ = 177^{+10}_{-6}$ km \, s$^{-1}$
(with the given uncertainty 
derived from the allowed $z_{\rm s}$ range).
This value is consistent with the limits imposed by the FJ relation.

\gl\ offers the possibility to investigate the mass distribution
of the very inner region
in an $L_*$ galaxy at $z \sim 0.5$.
Indeed, while there is growing evidence for
a picture in which
massive galaxies (i.e., above $\sim 10^{11.5} M_{\odot}$)
are well described by an isothermal mass profile and show
no structural evolution since $z\sim1$
(see, for instance, Treu~2007),
we still lack a large sample of $L_*$ galaxies
at $z\gtrsim0.5$
with a robust determination of the inner DM fraction
in order to probe mass assembly and galaxy evolution at lower mass scales.

We find that \gl\
(assuming $z_{\rm s} = 1.3$) has a projected DM fraction
$\sim 30$\% within the Einstein radius. 
Lower values of the source redshift require a larger
DM fraction in a non-spherical halo or a different IMF
(e.g., Fig.~\ref{fig:tre}).
By matching the measured mass-to-light ratio at the \RE\ with the one
expected from a double-component model formed by a NFW spherical halo
and a Hernquist spheroidal light distribution,
we derive the total dark-to-luminous mass ratio
$f_{\rm vir}  \equiv \massDM /M_{*} =  6 \pm 4$.
This value is close to the one obtained for galaxies
in the local Universe (N+05)
located in the transition regime between massive DM dominated galaxies
and lower-mass, DM deficient systems.

A firmer estimate of the inner DM and stellar content
in this $L_*$ galaxy requires a further observational effort
to determine the redshift of the lensed source.
A possible strategy
would involve, for instance, near-infrared spectroscopy
aimed  at detecting the H$_{\alpha}$ emission line.

\section*{Acknowledgments}

The authors thank J.M. Alcal\'a and M. Pannella for allowing them to
use the NTT spectrum of the lensing galaxy, C. Tortora for
discussions on the derivation of the stellar mass from spectroscopic
data, the Director of the Space Telescope Science Institute for granting
Director's Discretionary Time, and the referee for constructive 
comments.
GC and NRN acknowledge funding from EC through the FP6-European Reintegration Grants
MERG-CT-2005-029159 and MERG-CT-2005-014774, respectively.
JPK acknowledge support from CNRS.
OK acknowledges INTAS grant Ref.Nr.~05-109-4793.
NAG acknowledges support from grant HST-GO-10715.11-A.

\end{document}